# LINEAR AND NONLINEAR AC RESPONSE OF $MgB_2$ SUPERCONDUCTORS

M.I. Tsindlekht[*] and I. Felner
The Racah Institute of Physics, the Hebrew University of Jerusalem
91904 Jerusalem, Israel

We report the experimental study of linear and nonlinear ac properties of two types of samples of $MgB_2$ prepared by different technologies. A commercial sample (Alfa Aesar) demonstrates two stages of a broad phase transition, whereas home-made *metallic* sample have one sharp transition. The temperature dependence of the rectified signal shows two peaks for the commercial and one peak for the *metallic* samples. Two types of weak links (intergrain and intragrain) are responsible for the nonlinear response of these samples. The properties of these weak links are discussed.

The recent discovery [1] of superconductivty around $T_C$=39 K in the simple intermetallic compound $MgB_2$ is particularly surprising for many reasons. This material, available from common chemical suppliers, has been known and structurally characterized since the mid 1950`s. The hexagonal structure of $MgB_2$ is that of the well known $AlB_2$-type, which can be viewed as an intercalated graphite structure with full occupation of interstitial sites centered in a hexagonal prism consisting of B atoms. The simple structure of $MgB_2$ affects also the physical behavior of the samples, namely the weak superconducting intragrain and intergrain links. It is well known that for ceramic high-$T_c$ superconductor materials, the weak links produce both: (a) ac losses and (b) ac nonlinear harmonic generation [2, 3], whereas, for $MgB_2$ the weak links behave significantly different. For example, naturally-occurring grain boundaries are highly transparent to supercurrent, and therfore the critical current density ($J_C$) of the weak links is apparently comparable to $J_C$ of the grains [4].
The commercial bulk samples of $MgB_2$ are dark black powders. On the other hand, we have developed a method to prepare $MgB_2$ samples (with an excess of B) which exhibit a shiny *metallic* polycrystalline form. In this short communication we present an experimental study of the ac linear and ac nonlinear susceptibilty for the commercial and for our *metallic* $MgB_2$ materials.

The *metallic* MgB$_2$ material was prepared as follows. Stoichiometric ratios of the Mg and B elements (99.9% pure) in lump form were placed in a Ta tube. The Ta tube was then sealed in an evacuated quartz ampoule and heated to 950 C in a box furnace for two hours. The powder X-ray diffraction pattern was indexed to the well known hexagonal AlB$_2$-type unit cell of MgB$_2$, and the lattice parameters obtained are a=3.110 and c=3.519 , and are in excellent agreement with the data given in the literature. The pattern contained a few extra peaks (with intensity of less than 5%) which are due to an excess of B. The average diameter of the irregular *metallic* sample is ~3 mm. The commercial powder (Alfa Aesar, 98% pure) was pressed into a cylindrical shape with a diameter and height of ~3 mm. Linear ac measurements were performed using a standard two coil method at a frequency of $\Omega/2\pi = 733\,\text{Hz}$ and amplitude of excitation 0.03 Oe. The nonlinear experiments were carried out as follows. Each sample was exposed to an amplitude-modulated ac field $h(t) = h_0(1 + \alpha \cos \Omega t)\cos \omega t$, where the amplitude is $0 < h_0 < 0.9$ Oe, the modulation deep $\alpha \approx 1$, the carrier frequency $\omega/2\pi = 3.2\,\text{MHz}$; and modulation frequency $\Omega/2\pi = 733\,\text{Hz}$. When the sample is in a nonlinear state, it will produce on this excitation, magnetic moment oscillations not only at harmonics of the carrier and sideband frequencies, but with the frequency $\Omega$ as well as its harmonics. The signal on frequency $\Omega$, $A_\Omega$, was processed by a lock-in amplifier. Measurements of the rectified signal $A_\Omega$ were carried out by using home-made experimental set up [5].

The temperature dependence of the linear ac susceptibility $\chi'$ at zero dc magnetic applied field (H=0) for both commercial and *metallic* samples is shown in Fig.1(a). It is readily observed that the commercial sample exhibits two stages transition at T=27 K and at $T_c \approx 39\,\text{K}$. The width of both transitions is about 5 -7 K. We are aware to the fact the ac susceptibility studies reported by other group [6] do not show these two transitions, probably because of a difference in samples preparation. Apparently the *metallic* sample shows a one stage transition only at a temperature of $T_c \approx 39.6\,\text{K}$ and transition width of about 2.5 K. Fig. 1(b) shows the temperature dependence of the amplitude of the rectified signal $A_\Omega$ at H ~3 Oe. The commercial sample exhibits two peaks at T = 21.8 K and T = 37.4 K. The amplitude of both peaks is dc field dependent. At H= 15 Oe the first peak disappears, whereas the second peak is observable up to H ~50 Oe. Fig 1(b) shows that the *metallic* sample exhibits a single peak $A_\Omega$, at T = 38.3 K, and its amplitude is also field dependent. However, in contrast to the commercial sample, this peak was observable up to H ~ 1 kOe. The ac field amplitude dependence of the rectified signal $A_\Omega(h_0)$ up to $h_0$ =0.9 Oe, for both samples is quadratic over whole range of $h_0$ (see Fig. 2).

The weak links are the main source of the nonlinearity effects in our samples. The two peaks for the commercial sample are due to intergain and inragrain weak links. Intergrain weak links have low $T_C$ and $J_C$ values and this $J_C$ is highly affected by dc fields. A relatively low dc field (H=15 Oe) suppresses $J_C$ to zero which causes the first peak to disappear, whereas, the intragrain weak links are more stable with respect to dc fields, and the second peak is sustained up to about 50 Oe. On the other hand, the *metallic* sample has no intergrain weak links, and the intragrain weak links are highly transparent to a dc current [4]; therefore the rectified signal (Fig 1(b)) is observable up to H=1 kOe.

This research was supported by the Israel Academy of Science and Technology and by the Klachky Foundation for Superconductivity.

*email address: mtsindl@vms.huji.ac.il

Figure and captions

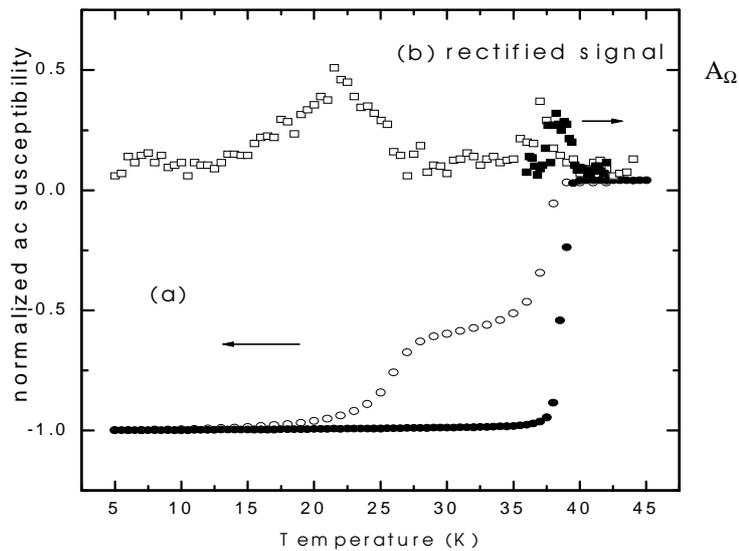

Fig.1
Fig1. (a)- Temperature dependence of the normalized linear susceptibility $\frac{\chi'(T)}{\chi'(5K)}$ for commercial (open circles) and *metallic* (filled circles) samples at zero dc field after ZFC.
(b)- Temperature dependence of the rectified signal for commercial (open squares) and *metallic* (filled square) samples at $h_0 = 0.6\,\text{Oe}$.

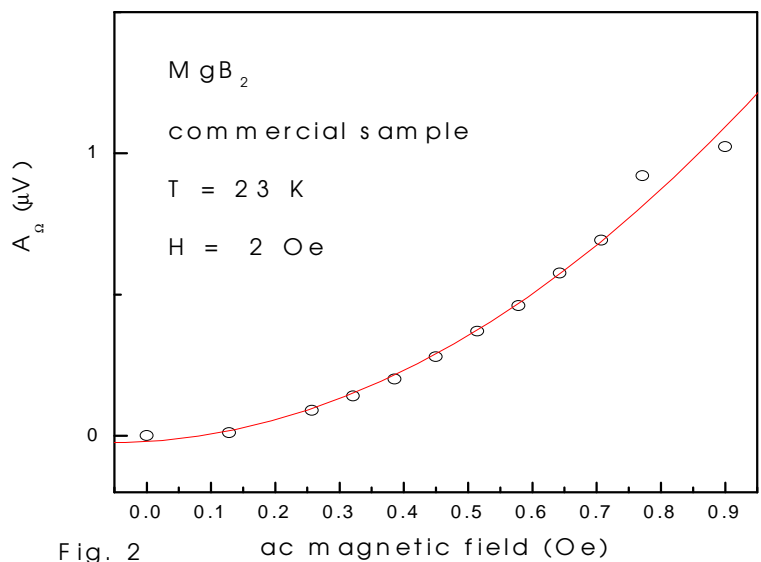

Fig. 2

Fig.2. Amplitude dependence of the rectified signal at T = 23 K and dc field about 2 Oe for the commercial sample.